\pdfoutput=1

\documentclass[3p,times,twocolumn]{elsarticle}

\usepackage{ecrc}

\volume{00}

\firstpage{1}

\journalname{Nuclear and Particle Physics Proceedings}

\runauth{G. Luparello for the ALICE Collaboration}

\jid{nppp}

\jnltitlelogo{Nuclear and Particle Physics Proceedings}

\usepackage{amssymb}

\usepackage[figuresright]{rotating}

\begin{document}

\begin{frontmatter}



\dochead{}

\title{ALICE mesurements of heavy-flavour production in pp and p-Pb collisions at the LHC}

\author{G. Luparello for the ALICE Collaboration}

\address{University of Trieste and INFN Trieste, Via A. Valerio 2, Trieste, Italy}

\begin{abstract}
The D meson yields as a function of charged-particle multiplicity in pp collisions at $\sqrt{s}=7$~TeV and in p-Pb collisions at $\sqrt{s_{\mathrm{NN}}}=5.02$~TeV are presented. The measurement of the yields of electrons from heavy-flavour hadron decays as a function of charged-particle multiplicity in p-Pb collisions at $\sqrt{s_{\mathrm{NN}}}=5.02$~TeV are shown as well. 
The measurement of azimuthal correlations of prompt D mesons and charged hadrons in pp collisions at $\sqrt{s}=7$~TeV and in p-Pb collisions at $\sqrt{s_{\mathrm{NN}}}=5.02$~TeV are also presented. The results are compared with expectations from models.
\end{abstract}


\end{frontmatter}


\section{Introduction}
\label{Intro}
The study of the production of hadrons containing charm and beauty quarks in proton-proton (pp) collisions at the LHC provides a way to test calculations based on perturbative QCD at high collision energy. In p-Pb collisions, heavy-flavour hadron production is sensitive to Cold Nuclear Matter (CNM) effects such as the modification of the parton distributions functions in the nucleus at small Bjorken-$x$ (see e.g.~\cite{Eskola:2009uj}), the parton energy loss in the initial stages of the collision via initial-state radiation~\cite{Vitev:2007ve} and the transverse momentum broadening due to soft scatterings of the partons in the incoming nucleus~\cite{Lev:1983hh}.

In addition to transverse momentum and rapidity differential distributions, measurements as a function of multiplicity and studies of angular correlations provide further constraints on the description of heavy-flavour production in pp and p-Pb collisions. The measurement of heavy-flavour production in pp collisions as a function of the charged-particle multiplicity could provide insights into the role of multi-parton interactions (MPI) and the interplay between hard and soft mechanisms in particle production. The multiplicity-differential measurements of heavy-flavour production in p-Pb collisions are sensitive to the dependence of CNM effects on the collision geometry and on the density of final-state particles. The measurement of azimuthal correlations of D mesons and charged particles in pp collisions provides a way to characterize charm production and fragmentation processes, while in p-Pb collisions they could give insights into possible collective effects in small systems.  

The excellent performance of the ALICE detector~\cite{Abelev:2014ffa} allows for open heavy-flavour measurements in several decay channels and in a wide rapidity range. This contribution focuses on the measurements at mid-rapidity where open heavy-flavour production is studied by means of fully reconstructed D mesons (in the decay channels $\mathrm {D^0} \rightarrow \mathrm {K^-} \pi^+$, $\mathrm{D^+} \rightarrow \mathrm{K^-} \pi^+ \pi^+$, $\mathrm{D^{*+}} \rightarrow \mathrm{D^0} \pi^+ \rightarrow \mathrm{K^-} \pi^+ \pi^-$, $\mathrm{D_{\mathrm s}^+} \rightarrow \phi  \pi^+ \rightarrow  \mathrm{K^-} \mathrm{K^+} \pi^+$ and their charge conjugates) and in the semi-electronic decay channels.

\section{Heavy-flavour production as a function of multiplicity}
\begin{figure}[!ht]
\begin{center}
\includegraphics[width=0.4\textwidth]{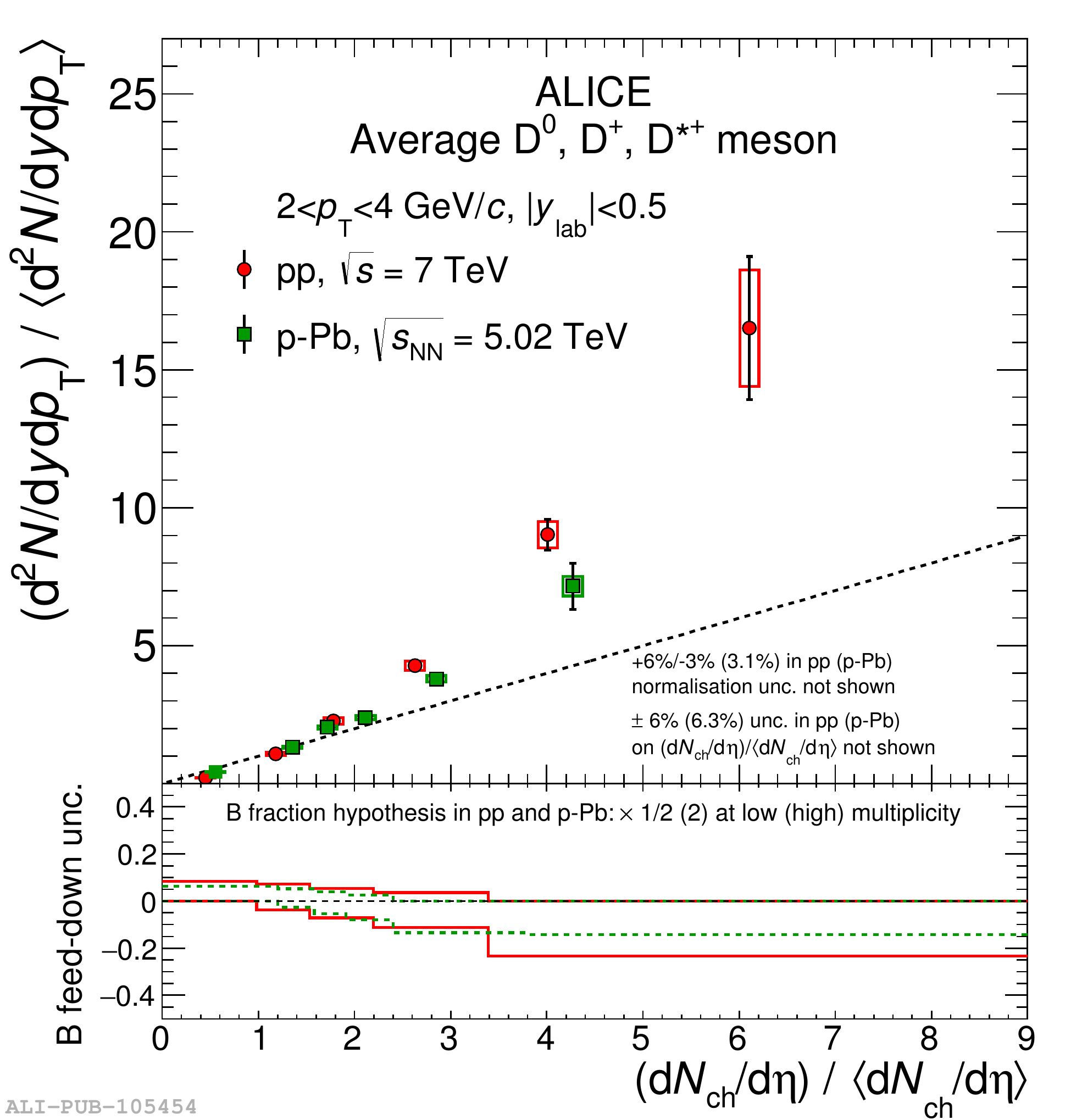}
\caption{Self-normalised yields of D mesons as a function of the charged-particle multiplicity estimated at mid-rapidity in pp (red) and p-Pb (green) collisions. }
\label{Fig:DvsMult}
\end{center}
\end{figure}
The study of heavy-flavour production as a function of the charged-particle multiplicity is presented via the self-normalised yields, i.e. the corrected per-event yield in a given multiplicity interval normalised to the multiplicity-integrated value. 
Multiplicity intervals at mid-rapidity are defined by measuring the number of tracklets (track segments joining a pair of particle hits in the two layers of the Silicon Pixel Detector and aligned with the reconstructed primary vertex) in the pseudorapidity interval $|\eta|<1$. The number of tracklets is converted to the charged-particle multiplicity at mid-rapidity ($ \mathrm{d}N_{\mathrm{ch}}/\mathrm{d}\eta $) by means of a Monte Carlo simulation. 
D$^{0}$, D$^{+}$ and D$^{*+}$ self-normalised yields are compatible with each other within uncertainties in the analyzed multiplicity and $p_{\mathrm{T}}$ intervals. The average D$^{0}$, D$^{+}$ and D$^{*+}$ self-normalised yields as a function of the charged-particle multiplicity at mid-rapidity in pp collisions and in p-Pb collisions are shown in Fig.~\ref{Fig:DvsMult}. A similar faster-than-linear increasing trend with charged-particle multiplicity is observed in the two colliding systems. In Ref.~\cite{Adam:2015ota} it is reported that the measurement in pp collisions is qualitatively described by model calculations taking into account the influence of the interactions between colour sources in the percolation model~\cite{Ferreiro:2012fb}, and the contribution of MPI, through the PYTHIA8~\cite{Sjostrand:2007gs} as well as the EPOS3 event generators~\cite{Werner:2013tya}.

\begin{figure*}[!ht]
\begin{center}
\includegraphics[width=0.6\textwidth]{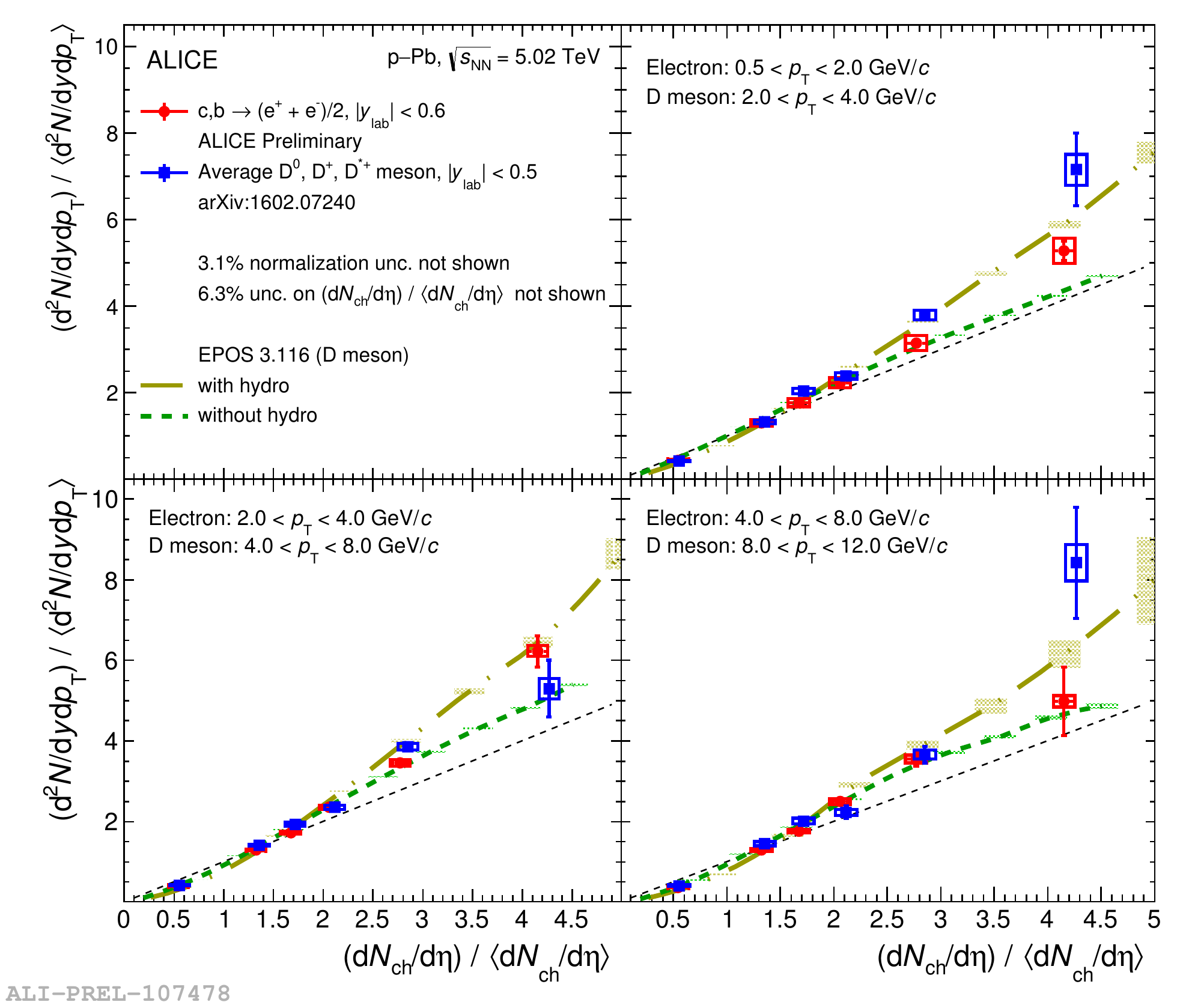}
\caption{Self-normalised yields of D mesons and of electrons from heavy-flavour hadron decays as a function of the charged-particle multiplicity estimated at mid-rapidity measured in p-Pb collisions. The results of EPOS3 model calculations~\cite{Werner:2013tya} are also shown. 
}
\label{Fig:DvsMult2}
\end{center}
\end{figure*}

Figure~\ref{Fig:DvsMult2} reports the self-normalised yields of electrons from semi-leptonic heavy-flavour hadron decays as a function of the charged-particle multiplicity estimated at mid-rapidity, compared with the D meson measurement. The different kinematic ranges for the two observables in
each panel approximately account for the effect of the decay kinematics. in each panel account for the decay kinematics. The increasing trend of D mesons and electrons from decay of heavy flavour hadrons is compatible within uncertainties. No dependence on the electron transverse momentum is observed also for $p_{\mathrm{T}}> \sim$4~GeV/c where a strong contribution from beauty-hadron decays has to be considered. The measurement gives a hint that the production mechanisms of charm and beauty as a function of the multiplicity are similar in p-Pb collisions. The results from model calculations using EPOS3 event generator~\cite{Werner:2013tya} for D mesons considering two approaches, i.e. with and without hydrodynamical evolution of the simulated initial conditions, are also shown. The calculations using a hydrodynamic evolution are in better agreement with the faster-than-linear increase of the D-meson measurement, suggesting that collective effects could play a role in particle production in high-multiplicity p-Pb collisions.

\section{Azimuthal correlations}
\begin{figure*}[!h]
\begin{center}
\includegraphics[width=0.6\textwidth]{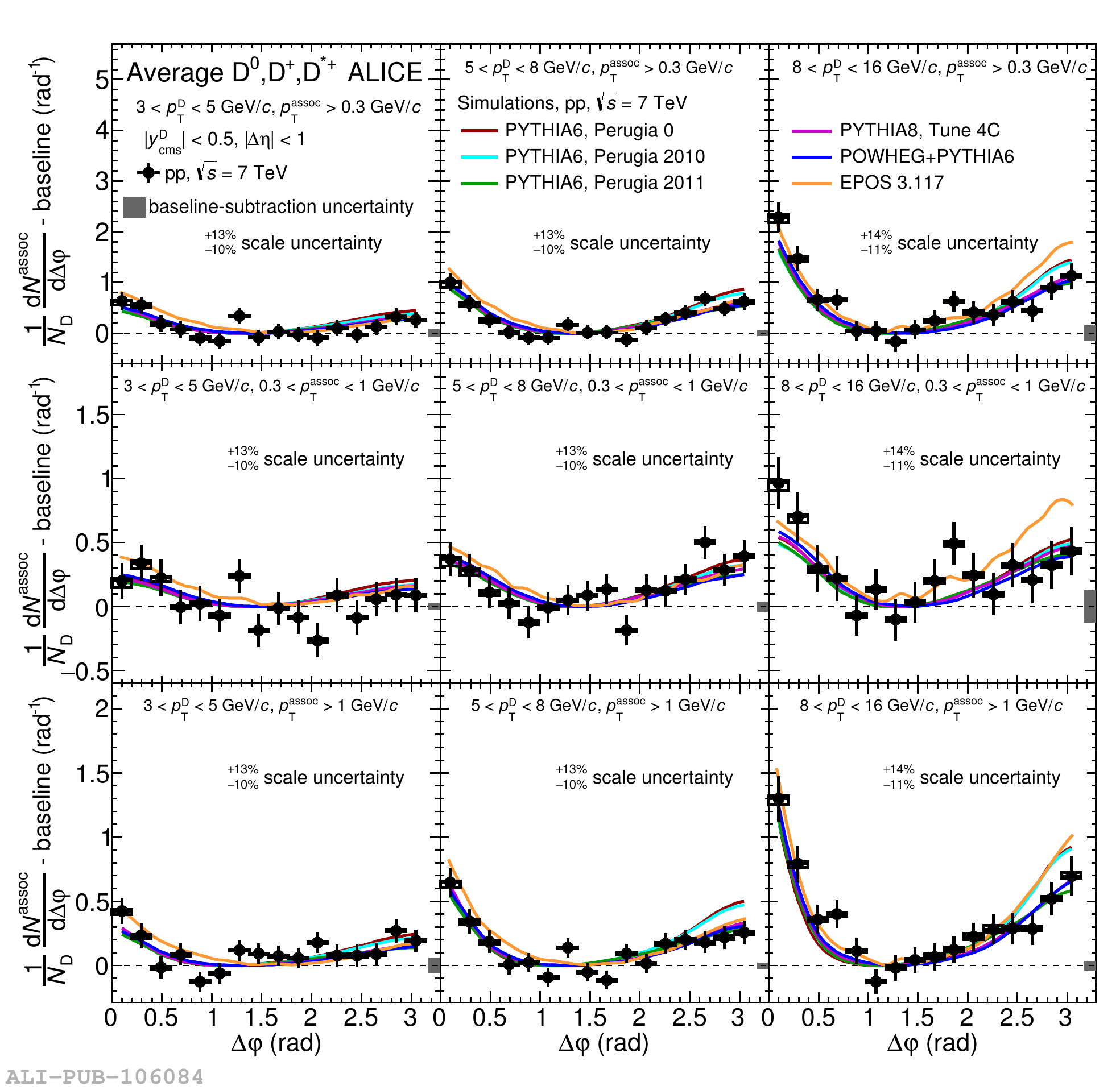}
\caption{Azimuthal-correlation distributions of prompt D mesons with charged particles measured in pp collisions at $\sqrt{s}=7$ TeV compared to expectations from Monte Carlo generators after the subtraction of the baseline.}
\label{Fig:Correlations}
\end{center}
\end{figure*}
\begin{figure*}[!hbt]
\begin{center}
\includegraphics[width=0.7\textwidth]{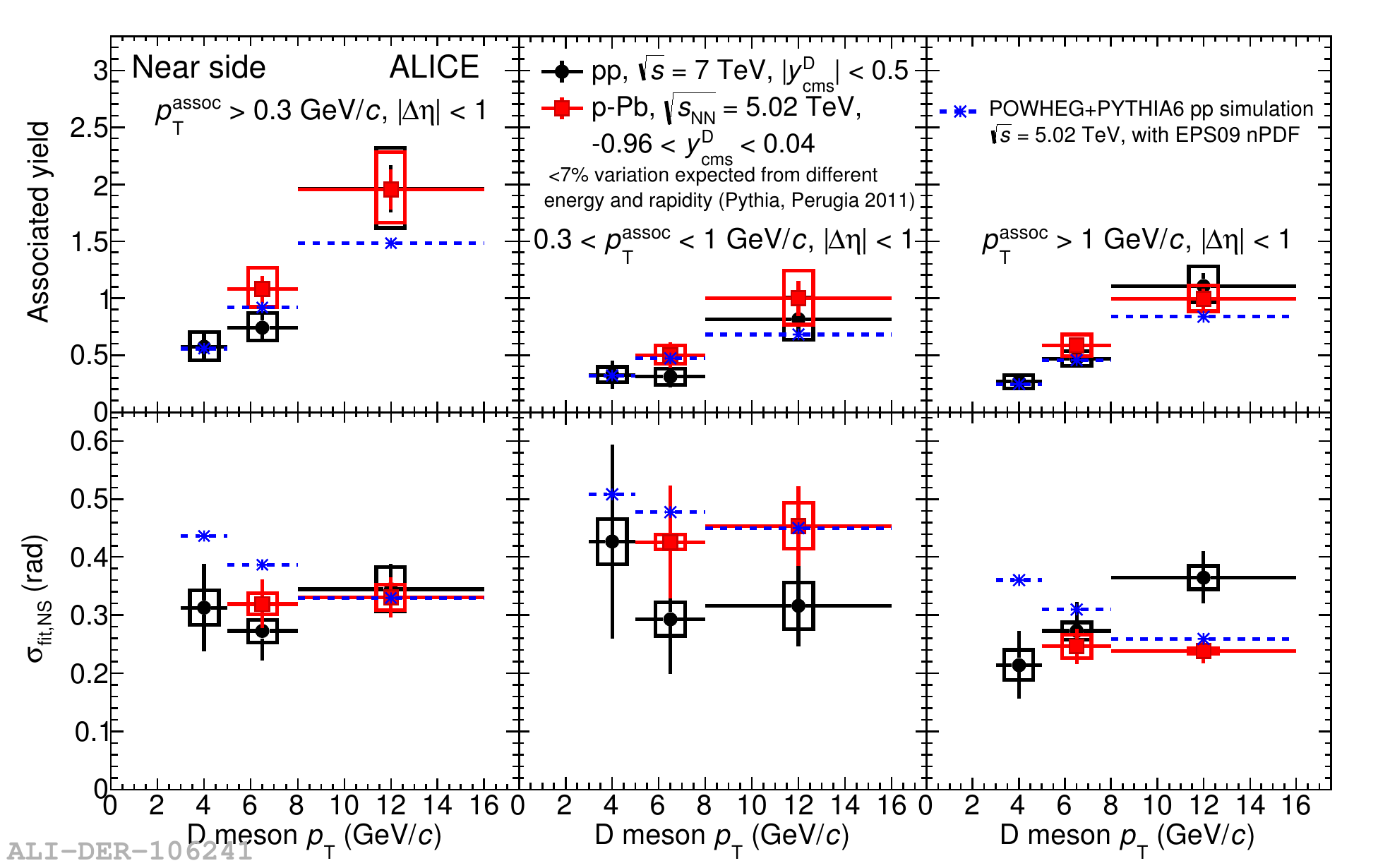}
\caption{Comparison of the near-side peak associated yield (top row) and peak width (bottom row) in pp and p-Pb collisions with expectations from POWHEG+PYTHIA6 Monte Carlo generator.}
\label{Fig:NearSide}
\end{center}
\end{figure*}
The azimuthal correlations are built by calculating the difference in $\Delta \eta$ and $\Delta \phi$ between a reconstructed D meson (trigger particle) and the primary charged particles (associated particles)~\cite{ALICE:2016clc}. Corrections for limited detector acceptance, spatial inhomogeneities, trigger and associated particle selection efficiency, as well as secondary track contamination are applied. The contribution due to the azimuthal correlations of D mesons from beauty-hadron decays and charged particles is obtained with PYTHIA simulations and subtracted. Different ranges of the D-meson ($3<p^{\mathrm{D}}_{\mathrm{T}}< 16$~GeV/c in pp, $5 < p^{\mathrm{D}}_{\mathrm{T}} < 16$~GeV/c in p-Pb) and associated particle $p_{\mathrm{T}}$ (starting from $p^{\mathrm{assoc}}_{\mathrm{T}} > 0.3$~GeV/c) are studied. 
The azimuthal correlation distribution is fitted with two Gaussian functions (one for the near-side peak centred at $\Delta \phi = 0$ and one for the away-side peak centred at $\Delta \phi = \pi$) on top of the baseline. The baseline is calculated from the region $\pi /4 < | \Delta \phi | < \pi /2$, and its variation upon redefinition of this interval is treated as a systematic uncertainty. 
Figure~\ref{Fig:Correlations} compares the baseline-subtracted D meson-charged particle azimuthal correlation distributions extracted in pp collisions with predictions by PYTHIA6~\cite{Sjostrand:2006za}, PYTHIA8~\cite{Sjostrand:2007gs} and POWHEG+PYTHIA6~\cite{Nason:2004rx,Frixione:2007vw} simulations, as a function of the D-meson $p_{\mathrm{T}}$, for different associated particle $p_{\mathrm{T}}$ ranges. The simulations reproduce the azimuthal correlation distributions within uncertainties, though a hint for a more pronounced near-side peak in data than in models is visible in the $8 < p^{\mathrm{D}}_{\mathrm{T}} < 16$ GeV/c range. The parameters extracted from the fit to the correlation distributions allow for a more quantitative comparison of the near-side peak properties. Figure~\ref{Fig:NearSide} compares the near-side associated yields and the near-side peak widths extracted in pp and p-Pb collisions. Compatible values of the near-side observables are obtained in pp and p-Pb collisions. No modifications of the near-side peaks due to cold nuclear matter effects are observed in p-Pb collisions with the current uncertainties. Predictions by POWHEG+PYTHIA6 simulations, including nuclear shadowing effects for the nucleon parton distribution functions, are also in agreement with the measurements.

\section{Conclusions}
The D mesons self-normalised yields are measured as a function of the charged-particle multiplicity estimated at mid-rapidity and the trend can be explained by model calculations with MPI in pp collisions~\cite{Adam:2015ota}. 
In p-Pb collisions a faster-than-linear increase of the self-normalised yields of D mesons and of electrons from heavy-flavour hadron decays as a function of the charged-particle multiplicity estimated at mid-rapidity is observed. The faster-than-linear increase suggests an interplay between MPIs and multiple binary nucleon-nucleon collisions in p-Pb interactions. 
The azimuthal correlation distributions of D mesons and charged particles, measured in pp and p-Pb collisions, as well as observables describing near-side peak properties, are in agreement between each other within uncertainties. These observables are also well described by PYTHIA and POWHEG+PYTHIA6 Monte Carlo simulations.

\nocite{*}



\begin{thebibliography}{00}


\bibitem{Eskola:2009uj} 
  K.~J.~Eskola, H.~Paukkunen and C.~A.~Salgado,
  JHEP {\bf 0904}, 065 (2009)

\bibitem{Vitev:2007ve}
  I.~Vitev,
  Phys.\ Rev.\ C {\bf 75} (2007) 064906

\bibitem{Lev:1983hh} 
  M.~Lev and B.~Petersson,
  Z.\ Phys.\ C {\bf 21}, 155 (1983).
  doi:10.1007/BF01648792

\bibitem{Abelev:2014ffa} 
  B.~B.~Abelev {\it et al.} [ALICE Collaboration],
  Int.\ J.\ Mod.\ Phys.\ A {\bf 29}, 1430044 (2014)
  

\bibitem{Adam:2015ota} 
  J.~Adam {\it et al.} [ALICE Collaboration],
  JHEP {\bf 1509}, 148 (2015)

\bibitem{Ferreiro:2012fb} 
  E.~G.~Ferreiro and C.~Pajares,
  Phys.\ Rev.\ C {\bf 86}, 034903 (2012)
  
\bibitem{Werner:2013tya} 
  K.~Werner, B.~Guiot, I.~Karpenko and T.~Pierog,
  Phys.\ Rev.\ C {\bf 89}, no. 6, 064903 (2014)
  
\bibitem{ALICE:2016clc} 
  J.~Adam {\it et al.} [ALICE Collaboration],
  arXiv:1605.06963 [nucl-ex].
  
\bibitem{Sjostrand:2006za} 
  T.~Sjostrand, S.~Mrenna and P.~Z.~Skands,
  JHEP {\bf 0605}, 026 (2006)
  
\bibitem{Sjostrand:2007gs} 
  T.~Sjostrand, S.~Mrenna and P.~Z.~Skands,
  Comput.\ Phys.\ Commun.\  {\bf 178}, 852 (2008)

\bibitem{Nason:2004rx} 
  P.~Nason,
  JHEP {\bf 0411}, 040 (2004)

\bibitem{Frixione:2007vw} 
  S.~Frixione, P.~Nason and C.~Oleari,
  JHEP {\bf 0711}, 070 (2007)


\end{thebibliography}
\end{document}